\documentclass[%
 reprint,
 twocolumn,
 nofootinbib,
 superscriptaddress,
 showpacs,
 preprintnumbers,
 amsmath,
 amssymb,
 aps,
 prd,
 floatfix,
]{revtex4-2}

\usepackage[british]{babel}

\usepackage[colorlinks=true, allcolors=blue]{hyperref}
\usepackage{braket}            %for braket notation
\usepackage{graphicx}          %Include figure files
\usepackage[utf8]{inputenc}    %For Bali paper references
\usepackage{dcolumn}           % Align table columns on decimal point
\newcolumntype{d}[1]{D{.}{.}{#1}}
\usepackage{color}             %for \textcolor
\usepackage{xcolor}
\usepackage[caption=false]{subfig}        %Subfigures
\usepackage{csquotes}        %enquote
\usepackage{mathtools}         %for \mathclap
\usepackage[normalem]{ulem}            %for \sout (strikethrough)
\usepackage{placeins}          %For the \FloatBarrier command - used manually
\usepackage{multirow}    % For a table with 'merged' cells
\usepackage{soul}

\usepackage{bm}% bold math
\definecolor{darkgreen}{rgb}{0,0.45,0}

\newcommand{\nn}{\nonumber}

\newcommand{\Tpc}{T_{\rm pc}}

\begin{document}

%https://www.overleaf.com/project/64e73bae6df644f07d59ad2f
\title{The curvature of the pseudo-critical line in the QCD phase diagram from mesonic lattice correlation functions}

 \author{Antonio Smecca}
 \affiliation{Centre for Quantum Fields and Gravity, Department of Physics, Swansea University, Swansea, SA2 8PP, United Kingdom}
\email{antonio.smecca@swansea.ac.uk (Corresponding author)}

 \author{Gert Aarts}
 \affiliation{Centre for Quantum Fields and Gravity, Department of Physics, Swansea University, Swansea, SA2 8PP, United Kingdom}
 
 \author{Chris Allton}
 \affiliation{Centre for Quantum Fields and Gravity, Department of Physics, Swansea University, Swansea, SA2 8PP, United Kingdom}

 \author{Ryan Bignell}
 \affiliation{School of Mathematics, Trinity College Dublin, Dublin, Ireland}

 \author{Benjamin J\"ager}
 \affiliation{Quantum Field Theory Center \& Danish IAS, Department of Mathematics and Computer Science \\
  University of Southern Denmark, 5230, Odense M, Denmark}
  
 \author{Seung-il Nam}
 \affiliation{Department of Physics, Pukyong National University (PKNU), Busan 48513, Korea}

 \author{Seyong Kim}
 \affiliation{Department of Physics, Sejong University, Seoul 143-747, Korea}

 \author{Jon-Ivar Skullerud}
 \affiliation{Department of Physics and Hamilton Institute, National University of Ireland Maynooth, County Kildare, Ireland}
 \affiliation{School of Mathematics, Trinity College Dublin, Dublin, Ireland}

 \author{Liang-Kai Wu}
 \affiliation{School of Physics and Electronic Engineering, Jiangsu University, Zhenjiang 212013, China 
and Key Laboratory of Quark and Lepton Physics (MOE), Central China Normal University, Wuhan 430079, China}

%\collaboration{\textsc{FASTSUM} Collaboration}%\noaffiliation

\date{\today}

\begin{abstract}
    In the QCD phase diagram, the dependence of the pseudo-critical temperature, $\Tpc$, on the baryon chemical potential, $\mu_B$, is of fundamental interest. The variation of $\Tpc$ with $\mu_B$ is normally captured by $\kappa$, the coefficient of the leading (quadratic) term of the polynomial expansion of $\Tpc$ with $\mu_B$.
    In this work, we present the first calculation of $\kappa$ using hadronic quantities.
    Simulating $N_f=2+1$ flavours of Wilson fermions on {\sc Fastsum} ensembles, we calculate the ${\cal O}(\mu_B^2)$ correction to mesonic correlation functions. By demanding degeneracy in the vector and axial-vector channels we obtain $\Tpc(\mu_B)$ and hence $\kappa$.
    While lacking a continuum extrapolation and being away from the physical point, our results are consistent with previous works using thermodynamic observables (renormalised chiral condensate, strange quark number susceptibility) from lattice QCD simulations with staggered fermions.
\end{abstract}

\maketitle

%%%%%%%%%%%%%%%%%%%%%%%%%%%%%%%%%%%%%%%%%%%%%%%%%%%%%%%
%%%%%%%%%%%%%%%%%%%%%%%%%%%%%%%%%%%%%%%%%%%%%%%%%%%%%%%

\section{Introduction}

A complete understanding of the phase diagram of quantum chromodynamics (QCD) in the temperature, $T$, and baryonic chemical potential, $\mu_B$, plane remains an open question in particle physics. The current understanding is that at $\mu_B=0$ a smooth crossover at a pseudo-critical temperature $\Tpc$ separates the low-$T$ phase, characterised by broken chiral symmetry, from the high-$T$ phase, in which chiral symmetry is restored~\cite{Karsch:2019mbv}.
This is supported by numerous lattice QCD studies~\cite{Borsanyi:2010bp,Bazavov:2011nk,HotQCD:2018pds,Borsanyi:2020fev}, the only non-perturbative \textit{ab initio} method for calculations in QCD.

Of particular interest is the determination of the pseudo-critical curve, $\Tpc(\mu_B)$,  in the $T-\mu_B$ plane.
Lattice studies have shown that $\Tpc$ decreases with $\mu_B$~\cite{Allton:2002zi,Gavai:2008zr,HotQCD:2018pds} and it has been argued that the crossover turns into a first-order phase transition for sufficiently large values of $\mu_B$~\cite{PhysRevLett.81.4816}. At the critical endpoint, separating the crossover and the first-order transition, the transition is expected to be second order.
Unfortunately, direct lattice simulations are restricted to $\mu_B=0$ due to the sign problem~\cite{deForcrand:2009zkb,Aarts:2015tyj}. This problem can be circumvented by simulations at imaginary $\mu_B$~\cite{Alford:1998sd,Lombardo:1999cz,deForcrand:2002hgr,deForcrand:2008vr,DElia:2002tig,DElia:2004ani,Azcoiti:2005tv,Chen:2004tb,Cea:2007vt,Wu:2006su,Nagata:2011yf,Cea:2009ba,Alexandru:2013uaa,Cea:2012ev,Alba:2017mqu,Vovchenko:2017xad} or by a Taylor expansion of physical observables  in $\mu_B/T$ around $\mu_B=0$~\cite{Allton:2002zi,Gavai:2003mf,Schaefer:2004en,Allton:2005gk}. 
Both methods are most accurate near $\mu_B=0$. The pseudo-critical curve can then be expressed as 
\begin{align}
  \frac{\Tpc(\mu_B)}{\Tpc(\mu_B=0)} = 1 - \kappa \left(\frac{\mu_B}{\Tpc(\mu_B=0)}\right)^2 + \mathcal{O}(\mu_B^4),
  \label{eq:kappa}
\end{align}
where $\kappa$ is the curvature, see the review~\cite{DElia:2018fjp}.

Thus far, observables used to determine $\Tpc(\mu_B)$ and calculate $\kappa$ are directly obtainable from the QCD partition function, namely the renormalised chiral condensate and the strange quark number susceptibility, 
see e.g.\ Refs.~\cite{Kaczmarek:2011zz,Endrodi:2011gv,Cea:2014xva,Bonati:2014rfa,Bonati:2015bha,Bellwied:2015rza,Cea:2015cya,Steinbrecher:2018phh,Bonati:2018nut,Borsanyi:2020fev} for studies with (partially) chirally symmetric actions at or close to the physical point and extrapolated to the continuum limit. 
In this work we introduce a new lattice approach based on hadronic physics.
We first expand mesonic correlation functions as a Taylor series in $\mu_B$ and note that in the presence of chiral symmetry a degeneracy is expected for correlation functions in the light vector and axial-vector meson channels.
By mapping out the temperature where this degeneracy occurs as a function of $\mu_B$, we determine the pseudo-critical curve, $\Tpc(\mu_B)$, and hence $\kappa$.

The paper is organised as follows: in Sec.~\ref{sec:methods} we introduce the lattice ensembles and the quantities used in the analysis, in Sec.~\ref{sec:results} we present the numerical results, in Sec.~\ref{sec:systematics} we discuss the systematic effects that could affect our calculation, and in Sec.~\ref{sec:conclusion} we summarise our conclusions.
Preliminary results for mesonic correlation functions at nonzero $\mu_B$ can be found in Ref.~\cite{Nikolaev:2020vll}.

%%%%%%%%%%%%%%%%%%%%%%%%%%%%%%%%%%%%%%%%%%%%%%%%%%%%%%%
%%%%%%%%%%%%%%%%%%%%%%%%%%%%%%%%%%%%%%%%%%%%%%%%%%%%%%%

\section{Methodology}\label{sec:methods}

In our investigation we used two sets of {\sc Fastsum} collaboration lattice ensembles with $\mu_B = 0$, labelled ``Generation 2'' and ``Generation 2L''.
These are generated using $N_f=2+1$ dynamical fermions, setting the up and down quarks to be degenerate and with an approximately physical strange quark. We employ anisotropic lattices with $a_{\tau}/a_s \ll 1$ to increase the number of data points in the temporal direction. The fermions and gauge fields are simulated using the $\mathcal{O}(a)$-improved Wilson fermion action with stout-smeared links and a Symanzik-improved gauge action respectively.
%To vary the temperature of our ensembles
We use a fixed-scale approach where the temperature $T = 1/(a_{\tau}N_{\tau})$ changes by varying $N_{\tau}$ with $a_{\tau}$ kept fixed. While the lattice spacings of the two ensembles are nearly the same, the Generation 2L pion mass is much lighter than in Generation 2, albeit still heavier than in nature.
%The mass of the $\Omega$ baryon was used to determine $a_{\tau}$ on each ensemble~\cite{Edwards:2012fx,Wilson:2019wfr}. Here, the Generation 2 value has been updated compared to our previous work~\cite{Aarts:2014nba}, using the analysis presented in Ref.~\cite{Edwards:2012fx} by the HadSpec collaboration.
Tables~\ref{tab:ensembles} and \ref{tab:temperatures} summarise the important features of the Generation 2 and Generation 2L ensembles.
Further details can be found in Refs.~\cite{Aarts:2014nba, Gen2zenodo} and \cite{Aarts:2022krz,Aarts:2023nax,Gen2Lzenodo} respectively.

\begin{table}[t]
  \begin{center}
    \begin{tabular}{|c||c|c|}
      \hline
      & Generation 2 & Generation 2L\\
      \hline
      $M_\pi$ [MeV] & $391(3)$ & $239(1)$\\
      $\Tpc$ [MeV] & $182(2)$ & $167(3)$\\
      $a_{\tau}$ [fm] & $0.03482(26)$ & $0.03246(7) $\\
%      $a_{\tau}$ [GeV] & $5.63(4)$ & $5.997(34)$\\
%      $\xi = a_s/a_{\tau}$ & $3.444(6)$ & $3.453(6)$\\
      $a_s$ [fm] & $0.11992(92)$ & $ 0.11208(31)$\\
     $N_s$ & $24$ & $32$\\
%      $m_{\pi}L$ & $5.63$ & $4.36$\\
      \hline
    \end{tabular}
  \end{center}
  \caption{Generation 2 and Generation 2L ensembles.
  The pseudo-critical temperature $\Tpc (\mu_B = 0)$ is determined via the inflection point of the renormalised chiral condensate $\langle \overline{\psi}\psi \rangle_R $.
  As expected, the value depends on the pion mass $M_\pi$.
  The temporal (spatial) lattice spacings are $a_{\tau(s)}$, while the number of spatial points is $N_s$. For each generation the mass of the $\Omega$ baryon was used to determine $a_{\tau}$~\cite{Edwards:2012fx,Wilson:2019wfr}. Here, the Generation 2 value has been updated compared to our previous work~\cite{Aarts:2014nba}, using the analysis presented in Ref.~\cite{Edwards:2012fx} by the HadSpec collaboration.
  Further details can be found in Refs.~\cite{Aarts:2014nba,Gen2zenodo} (Generation 2) and \cite{Aarts:2022krz,Aarts:2023nax,Gen2Lzenodo} (Generation 2L).
  }
  \label{tab:ensembles}
\end{table}

\begin{table}[t]
    \centering
    \begin{tabular}{|c||c | c | c || c | c | c | c |}
    \hline
    \multicolumn{8}{|c|}{Generation 2}\\
    \hline
    $N_{\tau}$ & $40$ & $36$ & $32$ & $28$ & $24$ & $20$ & $16$ \\
    $T$ $\mathrm{[MeV]}$ & $141$ & $156$ & $176$ & $201$ & $235$ & $281$ & $352$\\
    $T/\Tpc$ & $0.77$ & $0.86$ & $0.97$ & $1.10$ & $1.29$ & $1.54$ & $1.93$\\
    $N_{\rm cfg}$ & $501$ & $501$ & $1000$ & $1001$ & $1001$ & $1001$ & $1001$ \\
    \hline
    \end{tabular}
    \vspace{.4cm}
    \begin{tabular}{|c||c || c | c | c | c | c | c |}
    \hline
    \multicolumn{8}{|c|}{Generation 2L}\\
    \hline
    $N_{\tau}$ & $40$ & $36$ & $32$ & $28$ & $24$ & $20$ & $16$ \\
    $T$ $\mathrm{[MeV]}$ & $152$ & $169$ & $190$ & $217$ & $253$ & $304$ & $380$\\
    $T/\Tpc$ & $0.91$ & $1.01$ & $1.13$ & $1.29$ & $1.51$ & $1.82$ & $2.27$\\
    $N_{\rm cfg}$ & $1102$ & $1119$ & $1090$ & $1031$ & $1016$ & $1030$ & $1102$\\
    \hline
    \end{tabular}
    \caption{Number of time slices and temperatures related via $T = 1/(a_{\tau}N_{\tau})$, as well as $T/\Tpc$ and the number of lattice configurations in the Generation 2 and 2L ensembles.
    }
    \label{tab:temperatures}
\end{table}

We compute meson correlation functions projected to zero momentum, 
\begin{align}
  G(\tau) = \langle J_H(\tau)J_H^{\dagger}(0)\rangle \equiv \langle g(\tau) \rangle_B,
  \label{eq:meson}
\end{align}
where $J_H = \overline{\psi}\Gamma_H \psi$ is a meson operator in the isotriplet channel $H$, with $\Gamma_H$ the appropriate combination of $\gamma$ matrices to match the quantum numbers of the desired state,
 $g(\tau)\equiv \langle J_H(\tau)J_H^{\dagger}(0) \rangle_F$ is the
 correlator averaged over fermionic degrees of freedom only, and $\langle.\rangle_B$ represents an average over the bosonic (gauge) degrees of freedom.

Following Refs.~\cite{Nikolaev:2020vll,QCD-TARO:2001lhr,Pushkina:2004wa}, we expand the correlator to second order in  $\mu_q/T$, 
\begin{align}
 \nn
 G(\tau;\mu_q) & =  G(\tau;\mu_q)|_{\mu_q=0} + \frac{\mu_q}{T}T\frac{\partial G(\tau;\mu_q)}{\partial \mu_q}\Big|_{\mu_q=0} \\
 & + \frac{1}{2}\frac{\mu_q^2}{T^2}T^2\frac{\partial^2 G(\tau;\mu_q)}{\partial \mu_q^2}\Big|_{\mu_q=0} + \mathcal{O}\left(\frac{\mu_q^3}{T^3}\right),
 \label{eq:expansion}
\end{align}
where $\mu_q = \mu_u = \mu_d$ is the quark chemical potential for the two light flavours. The chemical potential for the strange quark, $\mu_s$, is set to zero, which is well justified \cite{Bonati:2014rfa,Steinbrecher:2018phh,DElia:2018fjp}. Note that the quark and baryonic chemical potentials are trivially related via $\mu_B = 3 \mu_q$. $T$ dependence is suppressed throughout.

One can show that odd derivatives with respect to $\mu_q$ vanish~\cite{QCD-TARO:2001lhr}, while the second derivative at $\mu_q=0$ is
\begin{align}
 &\frac{\partial^2 G(\tau;\mu_q)}{\partial \mu_q^2} \Big|_{\mu_q=0}  \equiv
  G''(\tau) \nn \\
  &\qquad = \Big\langle g''(\tau)
  + g \frac{\Delta''}{\Delta}\Big\rangle_B - \langle g \rangle_B \Big\langle \frac{\Delta''}{\Delta}\Big\rangle_B.
\end{align}
Here $\Delta=\det D[U,\mu_q]$ is the determinant for the two light quarks, the determinant for the strange quark is included (with $\mu_s=0$) but not written explicitly, 
$D$ is the Dirac operator, $U$ is the gauge link variable and the primes denote derivatives with respect to $\mu_q$. The full expression can be found in the appendix of Ref.~\cite{QCD-TARO:2001lhr}.
The numerically expensive part is the computation of the disconnected contribution,  which we evaluate using around
2000 random Gaussian noise sources \cite{Nikolaev:2020vll,TXL:1996zyh}. The correlators used in this work are presented in Appendix~\ref{app:Corrs}.

%%%%%%%%%%%%%%%%%%%%%%%%%%%%%%%%%%%%%%%%%%%%%%%%%%%%%%%
%%%%%%%%%%%%%%%%%%%%%%%%%%%%%%%%%%%%%%%%%%%%%%%%%%%%%%%

\section{Results}\label{sec:results}

From now on we focus on the vector and axial-vector correlators, which we denote here as $G_V$ and $G_A$, with $\Gamma_H = \gamma_i$ and $\gamma_5 \gamma_i$
respectively, summed over $i$, see Eq.~(\ref{eq:meson}). We consider these to second order, 
\begin{align}
   &G_{V/A} (\tau;\mu_q) = G_{V/A}(\tau) + \frac{1}{2}\mu_q^2G''_{V/A}(\tau).
\end{align}
To probe the restoration of chiral symmetry, we introduce the following ratio~\cite{Datta:2012fz}
\begin{align}
  R(\tau;\mu_q) = \frac{\tilde{G}_V(\tau;\mu_q)-\tilde{G}_A(\tau;\mu_q)}{\tilde{G}_V(\tau;\mu_q) + \tilde{G}_A(\tau;\mu_q)},
  \label{eq:R-ratio}
 \end{align}
 with
 $\tilde{G}_{V/A}(\tau;\mu_q) = G_{V/A}(\tau;\mu_q)/G_{V/A}(N_\tau/2;\mu_q)$,
i.e., both $\tilde{G}_{V/A}(\tau;\mu_q)$ are normalised to unity at the midpoint of the temporal lattice, $\tau = N_\tau/2$. Note that the wave function renormalisation $Z_{V/A}$ drops out of the ratios and that $R(\tau=N_\tau/2;\mu_q)$ is trivially zero. 

\begin{figure}[t]
  \centering
  \includegraphics[width=0.48\textwidth]{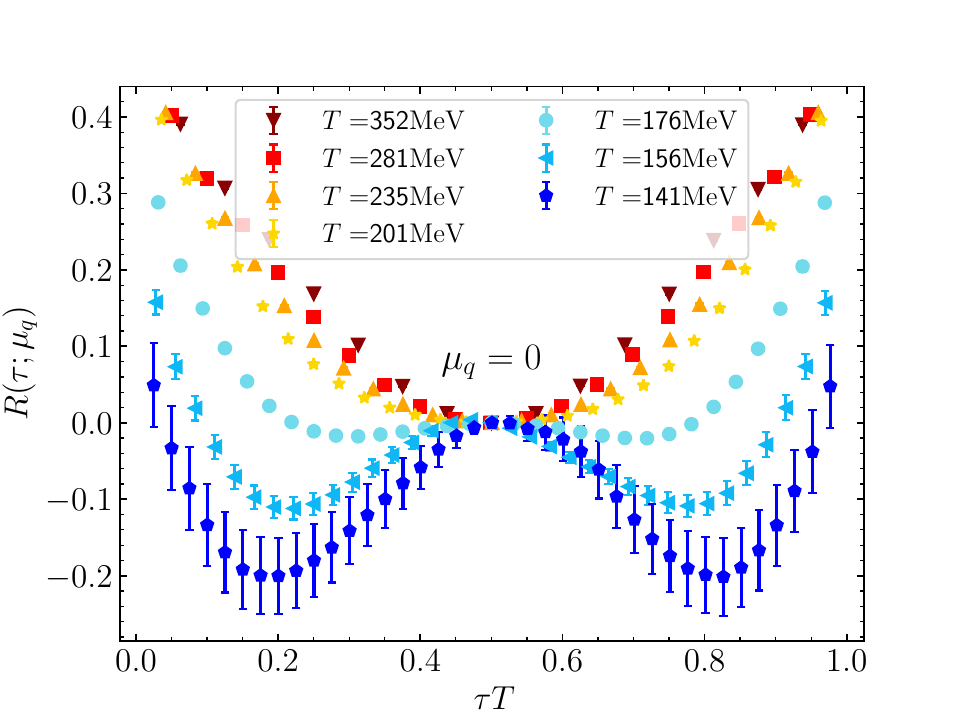}
  \includegraphics[width=0.48\textwidth]{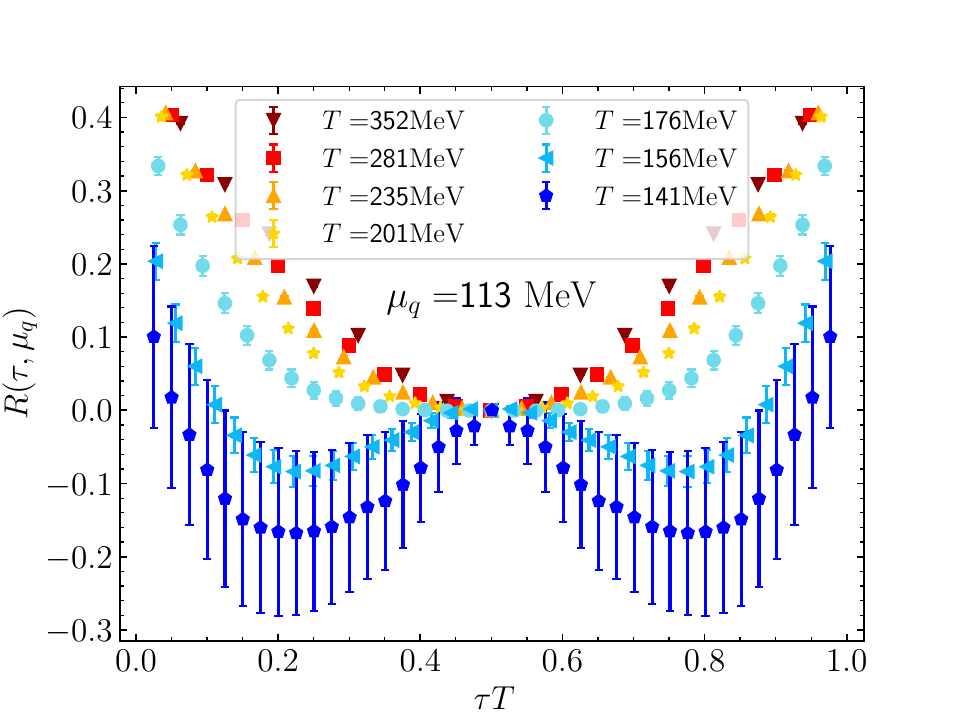}
  \caption{Generation 2 results for $R(\tau; \mu_q)$ as defined in Eq.~(\ref{eq:R-ratio}) for several temperatures at $\mu_q=0$ (above) and $\mu_q=113$ MeV (below).
  }
  \label{fig:R-ratio}
\end{figure}

\begin{figure}[t]
    \centering
    \includegraphics[width=0.48\textwidth]{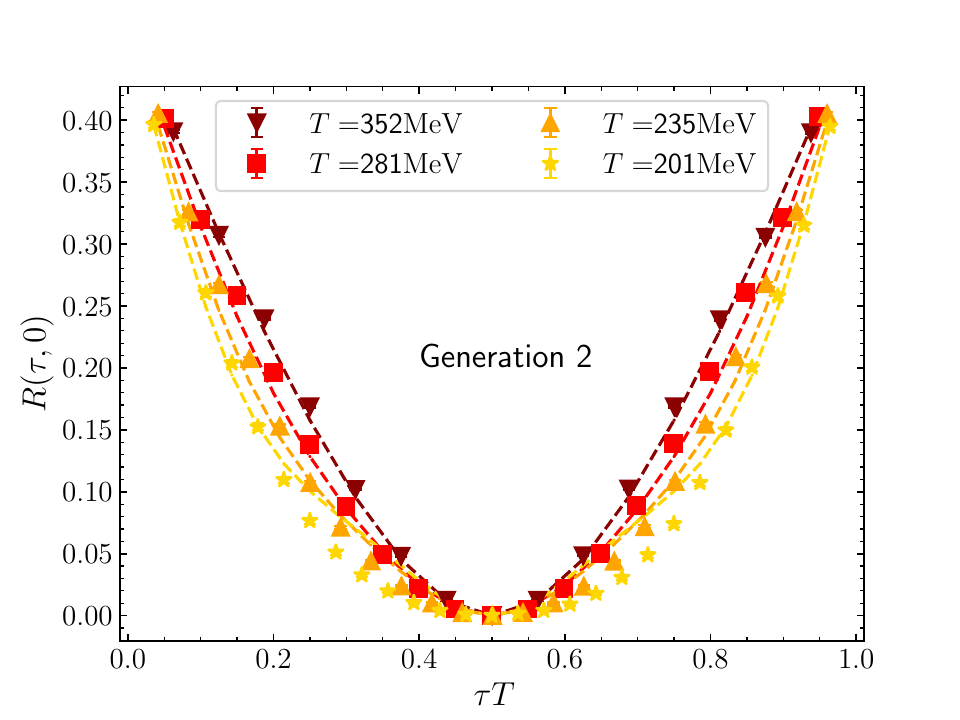}
    \caption{Comparison between $R(\tau; 0)$ on the Generation 2 ensembles in the high-temperature phase (symbols) and for non-interacting lattice fermions, using the same parameters and lattice geometry (dashed lines). 
        }
    \label{fig:R-model}
\end{figure}

In Fig.~\ref{fig:R-ratio} we present the ratio $R(\tau;\mu_q)$ for the Generation 2 ensembles at seven different temperatures at $\mu_q=0$ (above) and $\mu_q=113$ MeV (below). We concentrate on the behaviour near the centre of the lattice, $\tau T=1/2$. Consider first low temperatures in the confined phase. Assuming a single-pole Ansatz for the vector and axial-vector correlators, with dimensionless masses $\tilde m_{V/A}=m_{V/A}/T$, Eq.~(\ref{eq:R-ratio}) reads
\begin{align}
 R(\tau;0) & = \frac{\cosh(\tilde m_V x)-\cosh(\tilde m_Ax)}{\cosh(\tilde m_V x)+\cosh(\tilde m_Ax)} \nn \\
 & \approx \frac{1}{4T^2}\left(m_V^2-m_A^2\right)x^2 + {\cal O}(x^4),
\end{align}
 where $x= \tau T-1/2$ and the expansion is valid around the centre of the lattice. Indeed, given that the (ground state) vector meson is lighter than the axial-vector meson, we observe in Fig.~\ref{fig:R-ratio} an inverted parabola near the centre of the lattice at the lower temperatures. 
Away from the centre of the lattice, $R(\tau; \mu_q)$ is dominated by excited states and the short-distance behaviour of Wilson fermions, which breaks chiral symmetry explicitly and will be discussed below.

Considering medium effects in the hadronic phase,  the negative curvature is expected to depend on both the temperature and the chemical potential and to vanish when chiral symmetry is restored. Indeed, we observe a change in curvature as the temperature increases. 
Since we use Wilson fermions, we do not expect to see a complete degeneracy  --- $G_V(\tau)=G_A(\tau)$ and $R(\tau)=0$ for all $\tau$ --- even when chiral symmetry is restored. At high temperature, this can be analysed by calculating $R(\tau; \mu_q)$ with non-interacting lattice quarks, using the same (Wilson) discretisation and geometry as in the simulation \cite{Aarts:2005hg}. Results for $R(\tau; 0)$ obtained for free quarks and for  the Generation 2 ensembles in the high-temperature phase are compared in Fig.~\ref{fig:R-model}. The non-interacting results are shown by the dashed lines (even though they are computed at the same discrete set of lattice points). We observe excellent agreement at the highest temperatures, indicating that the quarks are weakly interacting. More importantly, this result shows that the non-degeneracy is explained by the Wilson fermion discretisation in the lattice action \cite{Aarts:2005hg}. 
As a side note, for lattice actions with (a remnant of) chiral symmetry, such as staggered quarks, full degeneracy is observed at higher temperatures, see $e.g.$\ Fig.~1 of Ref.~\cite{Aarts:2006wt} for the case of $G_{PS}(\tau)$ and $G_S(\tau)$.

\begin{figure}[t]
  \centering
  \includegraphics[width=0.48\textwidth]{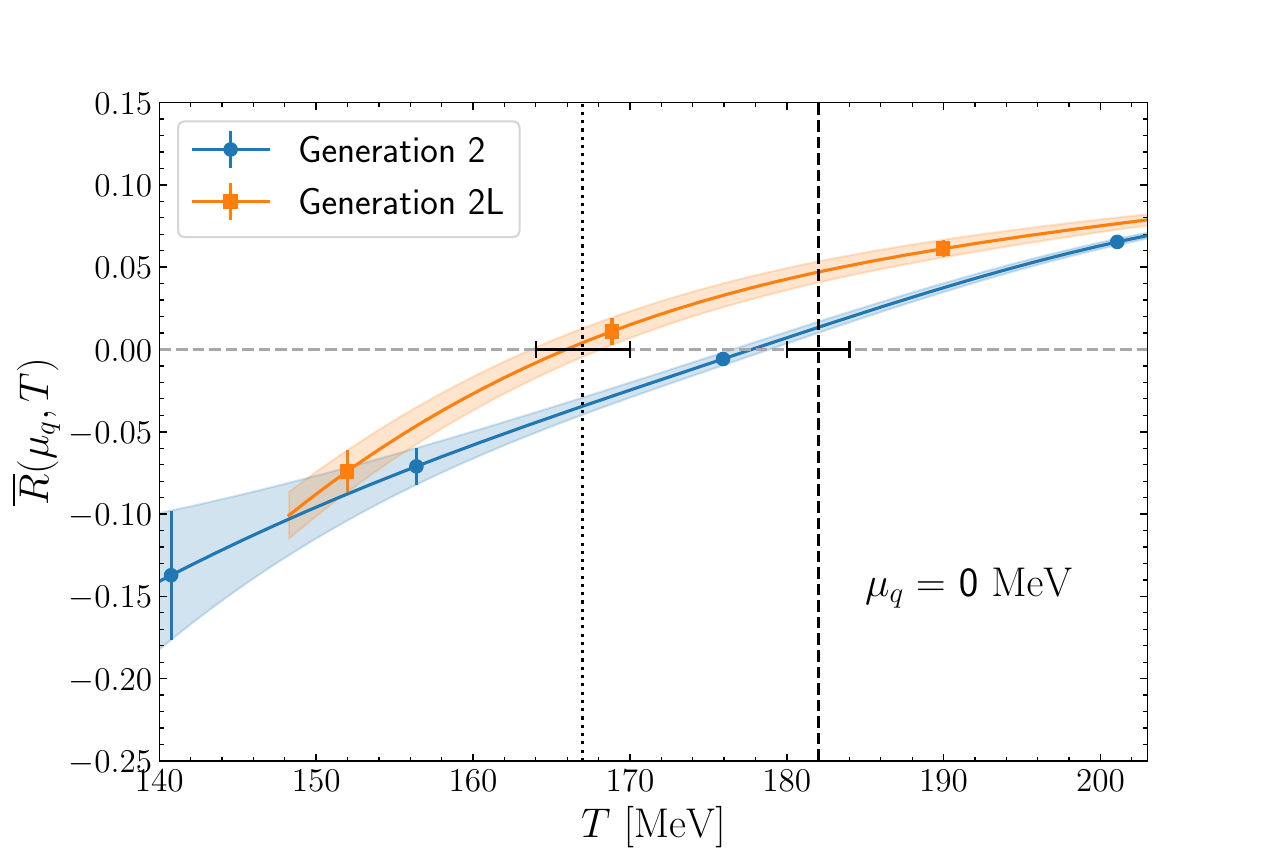}
  \includegraphics[width=0.48\textwidth]{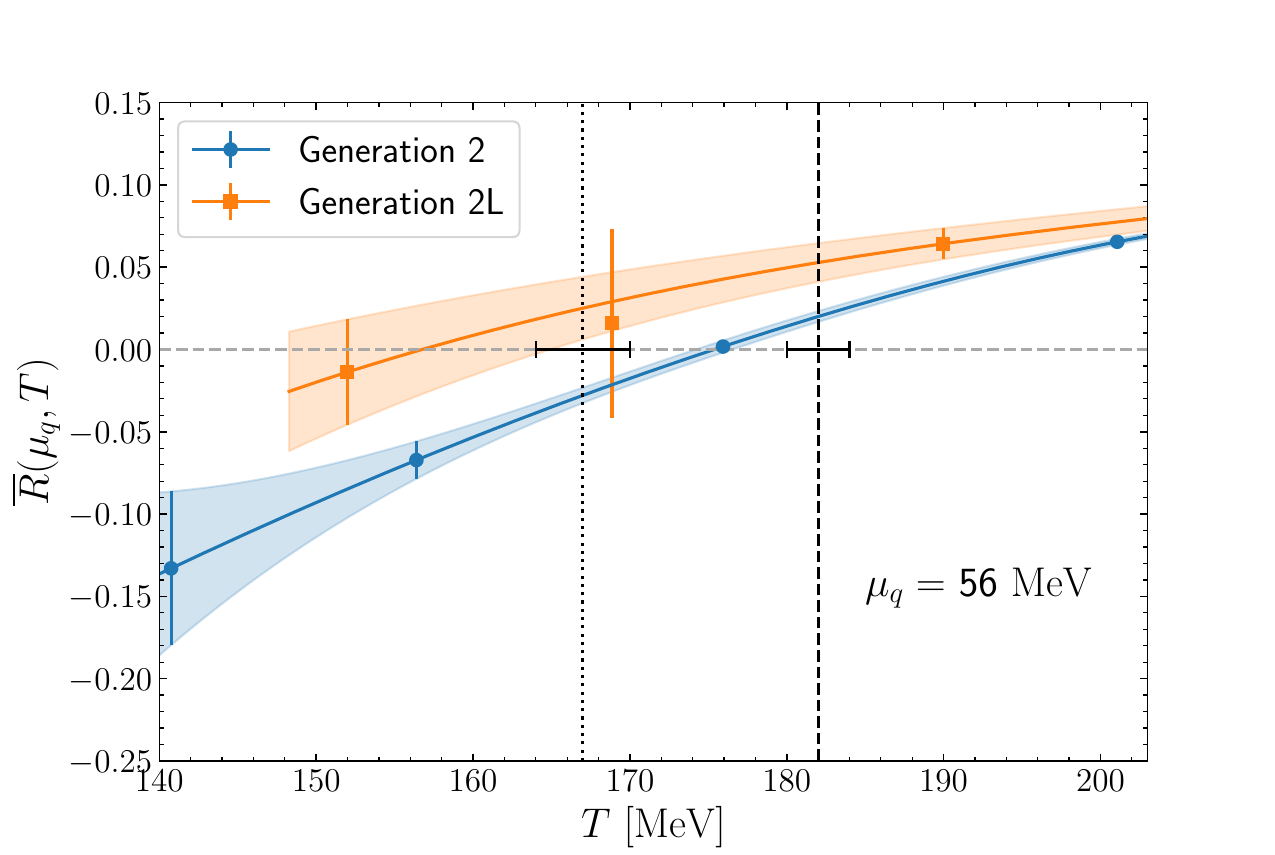}
  \caption{Averaged ratio $\overline{R}(\mu_q,T)$ as defined in Eq.~(\ref{eq:averaged_R}), plotted against the temperature, for $\mu_q=0$ (above) and $\mu_q=56$ MeV (below).
  Blue (orange) symbols correspond to Generation 2 (2L) and the blue (orange) curves show the corresponding interpolating curves obtained using cubic splines.
  The dashed (dotted) vertical lines correspond to $\Tpc(\mu_q=0)$ obtained from the renormalised chiral condensate for Generation 2 (2L), including the error estimate, see Table~\ref{tab:ensembles}.
  The $\mu_q$-dependent pseudo-critical temperature is determined by $\overline{R}(\mu_q, \Tpc)=0$ and decreases with $\mu_q$.
  }
  \label{fig:interpolation}
\end{figure}

The above discussion gives us confidence that we understand the behaviour of $R(\tau;\mu_q)$ both at low temperature, where $R(\tau;\mu_q)$ has a (local) maximum at $\tau T=1/2$, and at high temperature, where it has a minimum. In between these two regimes, by necessity there is a temperature where the curvature of $R(\tau,\mu_q)$ at the centre of the lattice vanishes. We use this point to define $\Tpc(\mu_q)$ in what follows. 
To quantitatively estimate the $\mu_q$ dependence of the pseudo-critical temperature, we define the time-averaged quantity~\cite{Aarts:2015mma,Aarts:2023nax},
\begin{align}
  \overline{R}(\mu_q,T) = \frac{\sum_{\tau=\tau_{\min}}^{N_{\tau}/2}R(\tau;\mu_q,T)/\sigma^2(\tau;\mu_q,T)}{\sum_{\tau=\tau_{\min}}^{N_{\tau}/2}1/\sigma^2(\tau;\mu_q,T)},
  \label{eq:averaged_R}
\end{align}
where $\sigma^2(\tau)$ is the variance at time $\tau$, and we have re-introduced the previously implicit dependence of $R$ on $T$. $R(\tau)$ is symmetric about $N_{\tau}/2$, and hence we use $N_{\tau}/2$ as the upper limit in the sum. The value of $\tau_{\min}$ is chosen so that $\tau_{\min}T \approx 0.2$ for all temperatures and for both Generation 2 and Generation 2L data. This value is selected to suppress unwanted effects due to the short-distance behaviour of Wilson fermions and excited states, see Fig.~\ref{fig:R-ratio}.
At fixed $\mu_q$, we then interpolate $\overline{R}(\mu_q,T)$ with $T$ using a cubic spline and find $\Tpc(\mu_q)$ from the $x$-intercept, i.e., $\overline{R}(\mu_q,\Tpc)=0$. We note that this method uses the correlation functions directly and does not require a fit.

Examples of this procedure are shown in Fig.~\ref{fig:interpolation} for both Generation 2 and 2L ensembles and at $\mu_q=0$ and $56$ MeV.
We also show the value of $\Tpc(0)$ found independently using the renormalised chiral condensate $\langle \overline{\psi}\psi \rangle_R$ \cite{Aarts:2020vyb}. 
As $\mu_q$ increases, the intercept moves to smaller values of $T$, reducing the pseudo-critical temperature. 
It is evident from the lower plot in Fig.~\ref{fig:interpolation} that the error bars on the Generation 2L data points are larger than for Generation 2, especially near $\Tpc$. This is mainly due to the smaller pion mass in the gauge ensembles. Moreover, the Generation 2L ensemble at  $T=169$ $\mathrm{MeV}$ sits, within errors, at the value of $\Tpc$ obtained via the renormalised chiral condensate. This leads to large fluctuations and adds no meaningful information. We hence have excluded this point from the analysis.

\begin{figure}[t]
  \centering
\includegraphics[width=0.48\textwidth]{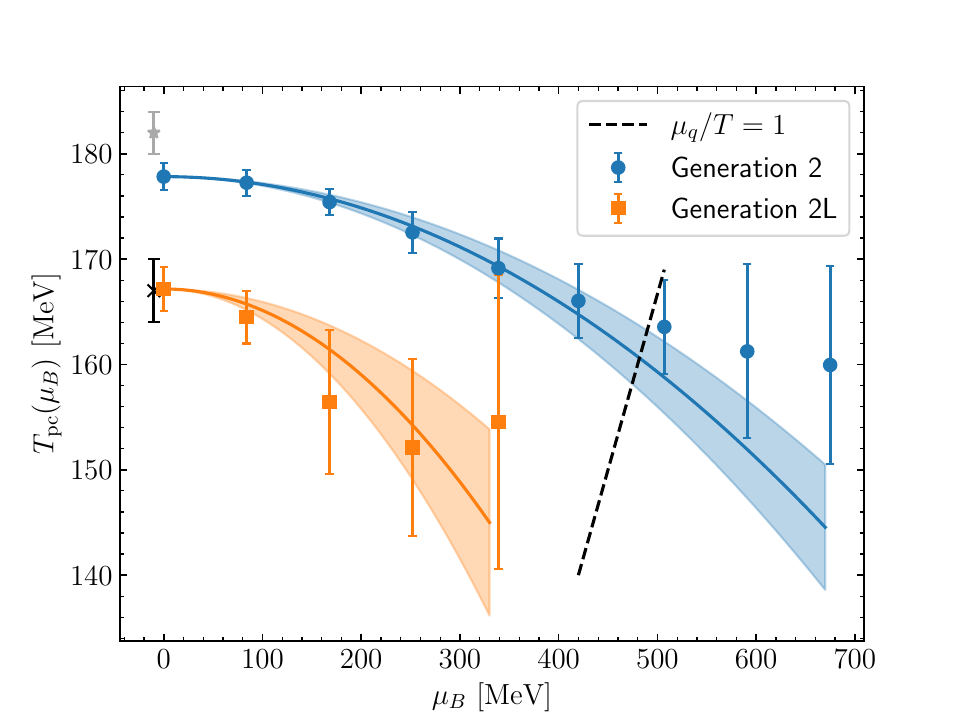}
  \caption{Pseudo-critical temperature as a function of $\mu_B$ from the Generation 2 (2L) ensembles shown by blue circles (orange squares).
  The blue (orange) curve is the fit of the data points to Eq.~(\ref{eq:kappa}) for Generation 2 (2L).
  The $T_{pc}(0)$ values obtained from the renormalised chiral condensate are shown as grey (black) crosses for Generation 2 (2L), see Table~\ref{tab:ensembles}.
  The dashed line is $\mu_q/T=1$ which sets a limit on the applicability of the Taylor expansion, Eq.~(\ref{eq:expansion}).
  }
  \label{fig:mu_fit}
\end{figure}

\begin{table}[t]
    \centering
    \begin{tabular}{|c | c || c | c |}
    \hline
     && Generation 2 & Generation 2L\\
    \hline
    $\mu_q$ $[\mathrm{MeV}]$ & $\mu_B$ $[\mathrm{MeV}]$ & $\Tpc(\mu_q)$ $\mathrm{[MeV]}$ & $\Tpc(\mu_q)$ $\mathrm{[MeV]}$ \\
    \hline 
    $0$  & $0$ & $177.8(1.2)(5.8)$ & $167.1(2.1)(3.1)$ \\
    $28$   &  $84$ & $177.2(1.2)(5.7)$ & $164.5(2.5)(3.1)$\\
    $56$   & $168$ & $175.4(1.2)(5.5)$ &  $156.5(6.8)(0.8)$\\
    $84$   & $252$ & $172.5(1.9)(3.9)$ & $152.1(8.4)(0.9)$\\
    $113$ & $339$ & $169.1(2.8)(1.5)$ & $155(14)(2.0)$\\
    $140$ & $420$ & $166.2(3.5)(0.6)$ & \\
    \hline
    \hline
    &$\kappa$ & $0.0131(23)(23)$ & $0.034(14)$ \\
    \hline
    \end{tabular}
    \caption{Pseudo-critical temperature $\Tpc(\mu_q)$ determined using Eq.~(\ref{eq:averaged_R}) for different values of $\mu_B = 3\mu_q$, for both the Generation 2 and 2L ensembles, including the values of $\kappa$ obtained from the fit to Eq.~(\ref{eq:kappa}). The results for $\Tpc(0)$ obtained from the renormalised chiral condensate are $\Tpc(0)=182(2)$ and $\Tpc(0)=167(3)$ for Generation 2 and 2L respectively, see Table~\ref{tab:ensembles}. The first error is statistical while the second is systematic. The value of $\kappa$ obtained using Generation 2L ensembles is dominated entirely by the statistical error.
    }
    \label{tab:Tc_mu}
\end{table}

The values for $\Tpc(\mu_q)$ obtained from this method for several values of $\mu_q$ are shown in Fig.~\ref{fig:mu_fit} and Table~\ref{tab:Tc_mu} for both Generation 2 and 2L. The gray and black dots in Fig.~\ref{fig:mu_fit} correspond to the value of $\Tpc(0)$ obtained via the renormalised chiral condensate, see Table~\ref{tab:ensembles}.
Also shown is the line $\mu_q/T=1$, which sets a limit on the applicability of the Taylor expansion, Eq.~(\ref{eq:expansion}).

\begin{figure}[t]
  \centering
  \includegraphics[width=0.48\textwidth]{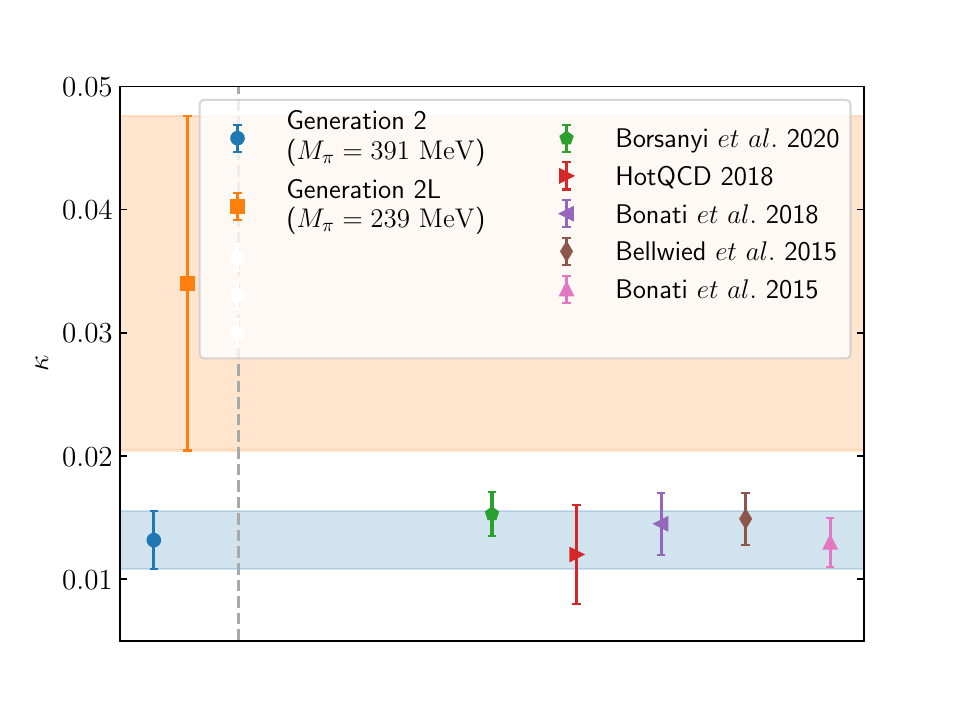}
  \caption{Final results for $\kappa$ obtained using Generation 2 (blue circle, $M_{\pi}=391(3)$ $\mathrm{MeV}$) and Generation 2L (orange square, $M_{\pi}=239(1)$ $\mathrm{MeV}$) compared with results from Refs.~\cite{Bonati:2014rfa,Bellwied:2015rza,Bonati:2018nut,HotQCD:2018pds,Borsanyi:2020fev}.
  }
  \label{fig:kappa}
\end{figure}

We fit the $\Tpc(\mu_q)$ values to Eq.~(\ref{eq:kappa}) to determine $\kappa$, obtaining the values shown in the bottom line of Table~\ref{tab:Tc_mu}. Our results for $\kappa$ together with those from other groups are plotted in Fig.~\ref{fig:kappa}.
For the Generation 2 ensembles, we note good agreement between our hadronic method and those using the renormalised chiral condensate. 
This agreement may be somewhat of a happy accident, as we have used one lattice spacing with unphysical light quarks, whereas the other results have been obtained in the continuum limit, with (partially) chirally symmetric actions, at or close to the physical point. 
Concerning Generation 2L, the lighter pion mass makes the calculation more challenging computationally which is reflected in a higher associated uncertainty. As stated, one of the ensembles sits very close to the transition, influencing the result strongly when included, while excluding it reduces the available data and hence increases the uncertainty.
%and lacking data when excluded.

Additionally, we note that, in principle, the curvatures in the $T-\mu_B$ plane could be different for different quantities since the transition is a crossover. If a critical endpoint exists, the transition lines would eventually coincide, however, the values of $\mu_B$ explored in this study are still too far away from any such putative critical endpoint to make a judgement on this.

%%%%%%%%%%%%%%%%%%%%%%%%%%%%%%%%%%%%%%%%%%%%%%%%%%%%%%%
%%%%%%%%%%%%%%%%%%%%%%%%%%%%%%%%%%%%%%%%%%%%%%%%%%%%%%%

\section{Systematics}\label{sec:systematics}

\begin{figure}[b]
  \centering
  \includegraphics[width=0.48\textwidth]{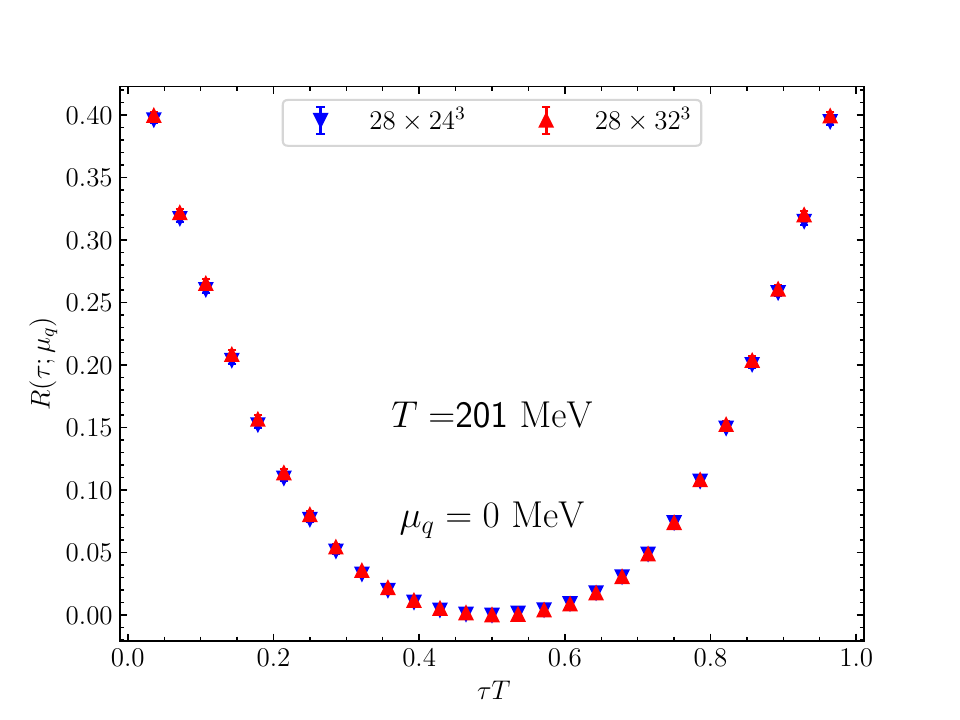}
  \includegraphics[width=0.48\textwidth]{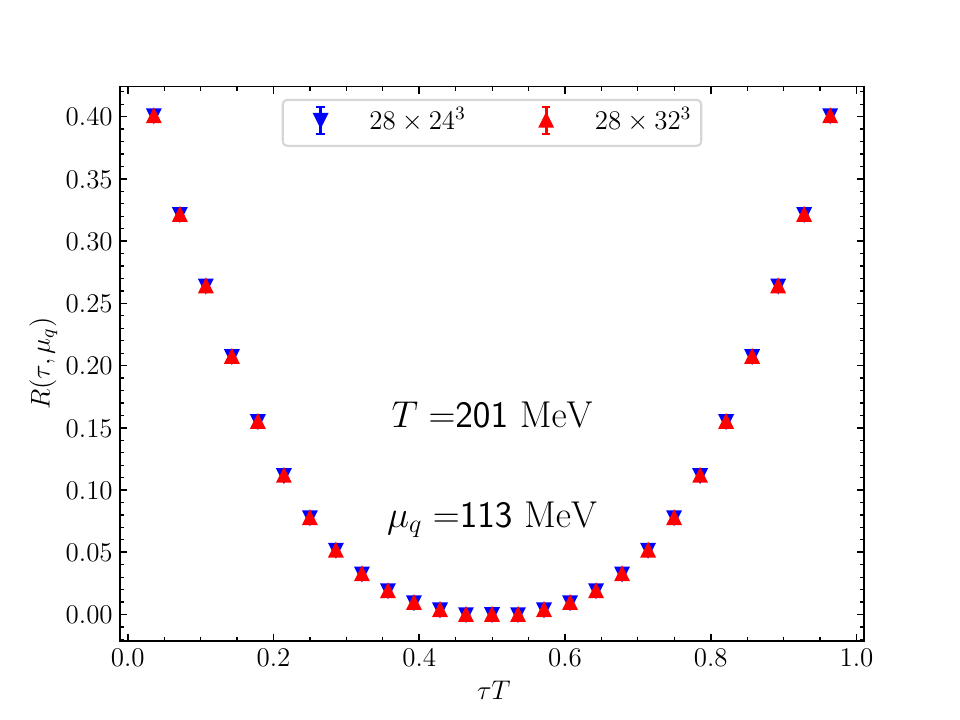}
  \caption{$R(\tau; \mu_q)$ ratio as defined in Eq.~\ref{eq:R-ratio} obtained using different volumes of the Generation 2 ensembles, for $\mu_q = 0$ (above) and $\mu_q = 113$ $\mathrm{MeV}$ (below) at fixed $T=201$ MeV.}
  \label{fig:volume-R_tau}
\end{figure}

\subsection{Finite volume effects}
\begin{figure}[t]
  \centering
  \includegraphics[width=0.48\textwidth]{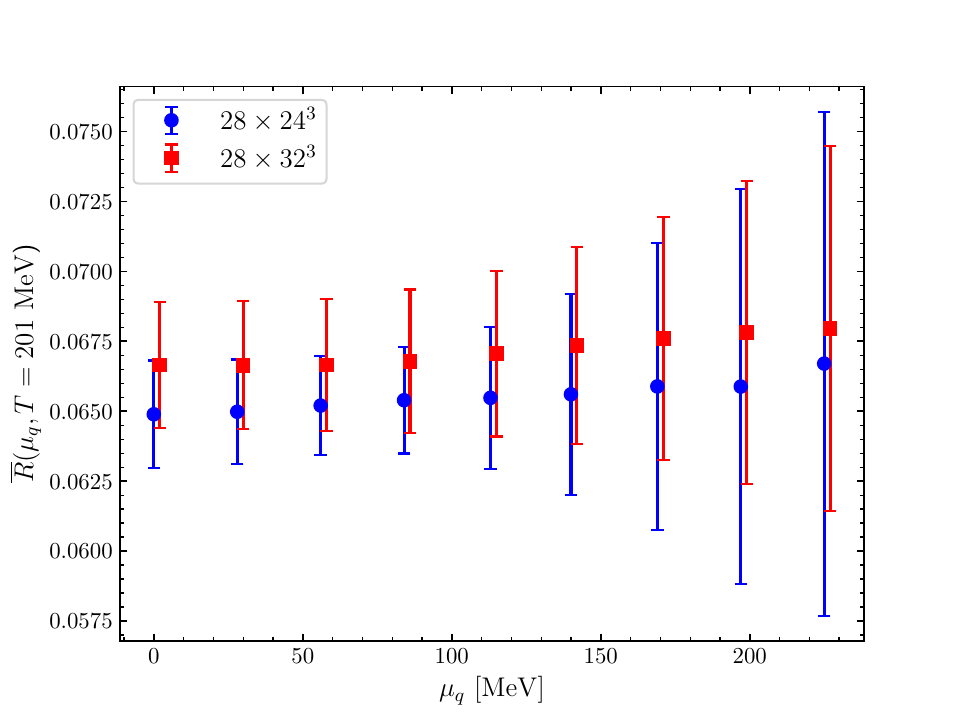}
  \caption{$\overline{R}(\mu_q, T)$ ratio as defined in Eq.~\ref{eq:averaged_R} at $T=201$ MeV  as a function of $\mu_q$. The blue circles and red squares correspond respectively to the spatial volume $N_s^3=24^3$ and $N_s^3=32^3$.}
  \label{fig:volume-R_tau_mu}
\end{figure}

In this section we discuss the systematics which could affect the result of our analysis. We start with finite-volume effects. To investigate the finite volume dependence, we produced correlation functions using a larger spatial volume for our Generation 2 ensemble at $T=201$ MeV ($N_{\tau}=28$), with the number of spatial lattice sites increased from $N_s=24$ to $N_s=32$. In Fig.~\ref{fig:volume-R_tau} we show results for $R(\tau; \mu_q)$, defined in Eq.~\ref{eq:R-ratio}, for the two different volumes  and for two different values of $\mu_q$.
%To check the size of finite volume effects, we compute the difference at $\mu_q = 0$ of $R(\tau)$ for the two volumes considered here. The results are shown in Fig.~\ref{fig:volume-difference} from which it is clear that the difference is always compatible with zero within the errorbars.
The negligible effect of increasing the volume is confirmed also on the averaged quantity $\overline{R}$, shown in Fig.~\ref{fig:volume-R_tau_mu}. The effect of changing the spatial volume gives results that are compatible well within the quoted statistical error for all $\mu_q$ values considered and hence we do not include it in the final error budget.

\subsection{Changing $\tau_{\min}$}
\begin{figure}
  \centering
  \includegraphics[width=0.48\textwidth]{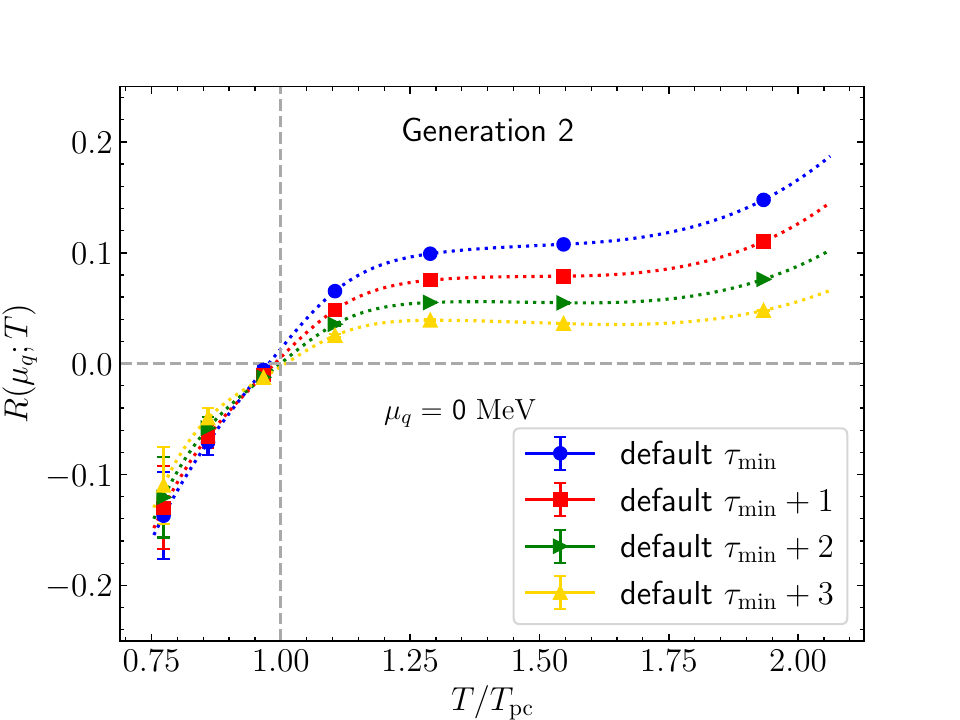}
  \includegraphics[width=0.48\textwidth]{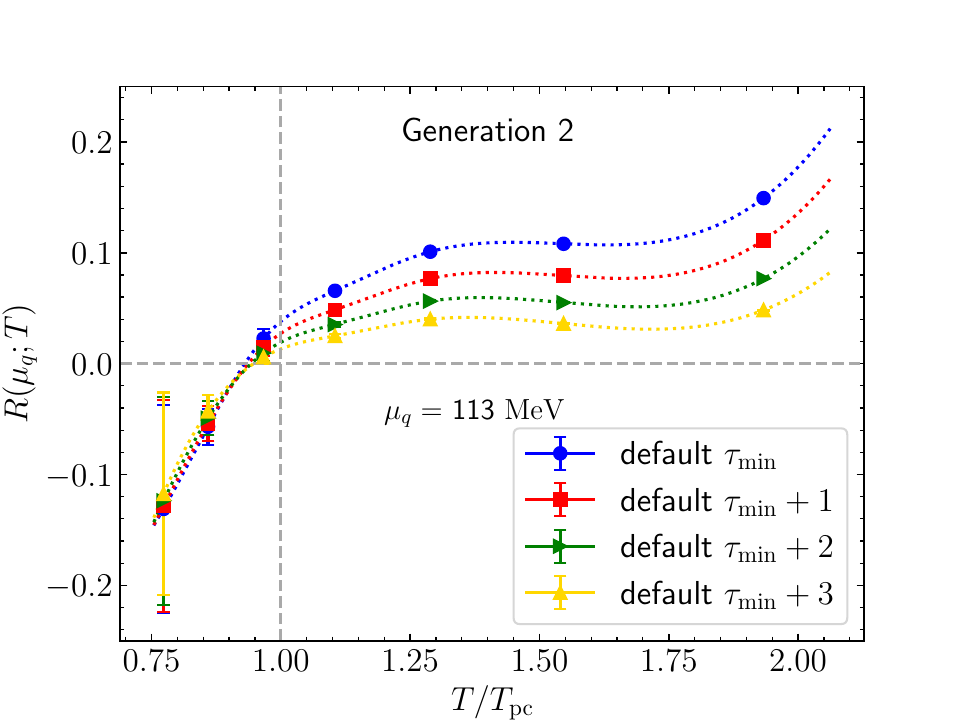}
  \caption{Generation 2 results for $\overline{R}(\mu_q,T)$ as a function of the temperature and for different choices of $\tau_{\min}$. The top and bottom plots show respectively the results at $\mu_q=0$ and $\mu_q=113$ $\mathrm{MeV}$. The vertical gray line shows the value of $T_{\mathrm{pc}}$ determined using the renormalised chiral condensate.}
  \label{fig:tmin-interpolate}
\end{figure}

We now turn our attention to the choice of $\tau_{\min}$. The default choices for $\tau_{\min}/a_{\tau}$ are $[3, 4, 4, 5, 6, 7, 7]$ and $[2, 3, 3, 4, 5, 6, 7] $ for Generation 2 and Generation 2L respectively, where the values are ordered from the smallest $N_{\tau}$ to the largest, $i.e.$ $N_{\tau} = [16,20,24,28,32,36,40]$. These values upon multiplication with the temperature give a value which corresponds roughly to $\tau_{\min}T \approx 0.2$. Below we explore different cases in which we increase the value of $\tau_{\min}$ up to $\tau_{\min}^{\mathrm{default}}+3$ (we put $a_\tau=1$ temporarily for notational convenience). Fig.~\ref{fig:tmin-interpolate} shows the value of $\overline{R}(\mu_q,T)$ for different choices of $\tau_{\min}$.
Below and up to $T_{\mathrm{pc}}$ the results are mostly compatible within the statistical errors. This is the main region of interest  since we determine the degeneracy of the vector and axial-vector channel at the point in which $\overline{R}$ crosses zero. 
As the temperature increases, the ratio becomes successively more sensitive to short-distance lattice artifacts of Wilson quarks, due to the choice of $\tau_{\rm min}$.

To compute the systematic error coming from varying $\tau_{\min}$ we use the following formula~\cite{ExtendedTwistedMassCollaborationETMC:2022sta}:
\begin{align}
  \Delta_{\mathrm{sys}} = \mathrm{max}\Bigg[ \Delta_{\mathrm{tot}} \mathcal{P}_{\mathrm{sys}}
   \, \mathrm{erf} \Big( \frac{\mathcal{P}_{\mathrm{sys}}}{\sqrt{2}} \Big) \Bigg],
\end{align}
with 
$\Delta_{\mathrm{tot}} = \sqrt{(\Delta T_{\mathrm{pc}}^{\mathrm{default}})^2 + (\Delta T_{\mathrm{pc}}^{\tau_{\min}})^2 }$ 
and 
$\mathcal{P}_{\mathrm{sys}} = |T_{\mathrm{pc}}^{\mathrm{default}} - T_{\mathrm{pc}}^{\tau_{\min}}|/ \Delta_{\mathrm{tot}}$. 
In the expression above, $\Delta T_{\mathrm{pc}}^{\mathrm{default}}$ and $\Delta T_{\mathrm{pc}}^{\tau_{\min}}$ are respectively the statistical error of the pseudo-critical temperature obtained using the default settings and using the value of $\tau_{\min}$ which shows the largest discrepancy from the default value. The pull variable $\mathcal{P}_{\mathrm{sys}}$ indicates how significant the systematic error is over the statistical error.

\begin{figure}[t]
  \centering
  \includegraphics[width=0.48\textwidth]{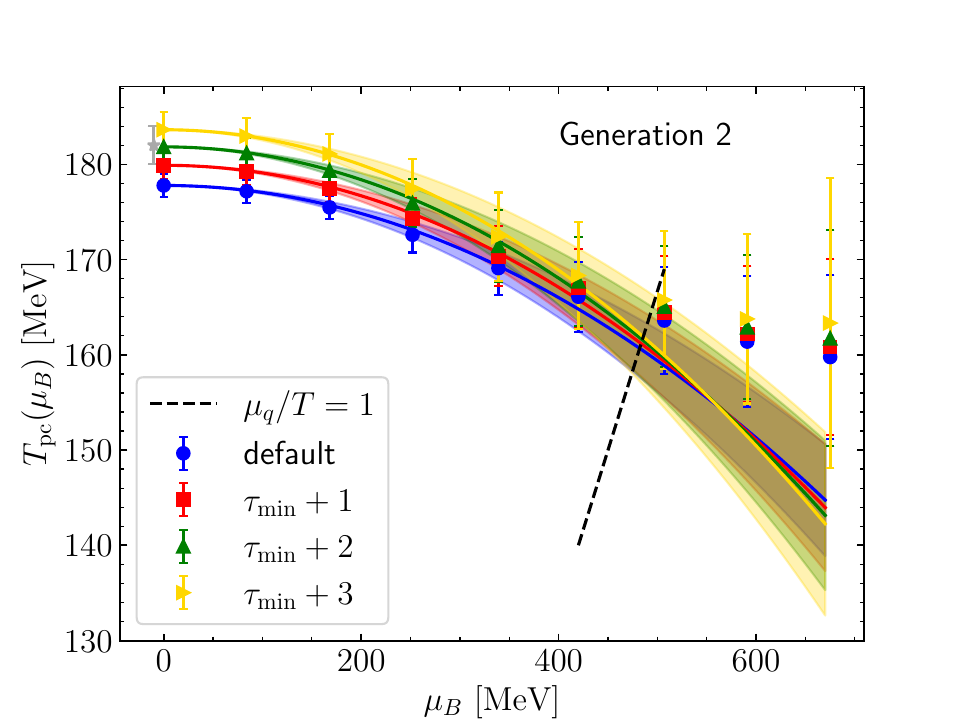}
  \caption{The pseudo-critical temperature as a function of $\mu_B$ for different choices of $\tau_{\min}$, from the Generation 2 ensembles. Similar results are obtained with Generation 2L ensembles.}
  \label{fig:tmin-Tc_mu}
\end{figure}

We investigated the systematic effect of varying $\tau_{\min}$ also on our fits of the pseudo-critical line to extract the value of $\kappa$. In Fig.~\ref{fig:tmin-Tc_mu} we show the fits for the Generation 2 ensembles and in Table~\ref{tab:tmin-kappa} we report the corresponding results for $\kappa$. Similar results are obtained with Generation 2L ensembles, where the final result for $\kappa$ is entirely dominated by statistical errors.

%\begin{figure}[t]
%  \centering
%  \includegraphics[width=0.48\textwidth]{tmin_final_kappa.pdf}
%  %\includegraphics[width=0.48\textwidth]{final_kappa_2L.pdf}
%  \caption{Values of $\kappa$ extracted by fitting the results of $T_{\mathrm{pc}}$ obtained for different values of $\tau_{\min}$. For Generation 2L we only show the results for $\kappa$ obtained from a fit with a $\chi^2/\mathrm{d.o.f} < 2$.}
%  \label{fig:tmin-kappa}
%\end{figure}

\begin{table}[t]
  \centering
  \begin{tabular}{| l | c |}
    \hline
    \;\;\;\;\;\;\;$\tau_{\min}$ & $\kappa$\\
    \hline
    default $\tau_{\min}$ & $0.0131(23)$\\
    default $\tau_{\min}+1$ & $0.0144(27)$\\
    default $\tau_{\min}+2$ & $0.0156(32)$\\
    default $\tau_{\min}+3$ & $0.0169(39)$\\
    \hline
  \end{tabular}
  \caption{Values of $\kappa$ extracted by fitting the results of $T_{\mathrm{pc}}$ obtained for different values of $\tau_{\min}$ with Generation 2 ensembles.}
  \label{tab:tmin-kappa}    
\end{table}
Even though the determination of $T_{\mathrm{pc}}$ is affected by the choice of $\tau_{\min}$, the value of $\kappa$ that we extract from our fits is rather stable, indicating that the $\tau_{\min}$ dependence is under control. 
%By increasing the value of $\tau_{\min}$ we are effectively restricting the range of points that we use in the average of $\overline{R}$, which for the hottest temperature reduces to just three points or less. It comes with no surprise that this is the range most affected by the change in $\tau_{\min}$. However, the points above $T_{\mathrm{pc}}$ have a marginal effect in the determination of $T_{\mathrm{pc}}(\mu_q)$ from the analysis presented above.
The systematic error on the values of $T_{\mathrm{pc}}$ obtained through our analysis is reported in Table~\ref{tab:Tc_mu} and the systematic error for $\kappa$ is included in Fig.~\ref{fig:kappa}.

%%%%%%%%%%%%%%%%%%%%%%%%%%%%%%%%%%%%%%%%%%%%%%%%%%%%%%%
%%%%%%%%%%%%%%%%%%%%%%%%%%%%%%%%%%%%%%%%%%%%%%%%%%%%%%%

\section{Conclusion}\label{sec:conclusion}

In this work we presented a novel approach to study the curvature of the pseudo-critical temperature, $\Tpc(\mu_B)$, with baryon chemical potential using meson correlation function obtained from lattice QCD.
We defined $\Tpc(\mu_B)$ via the degeneracy in the vector and axial-vector channels. The method employs the correlation functions directly and does not require fits.

We compared our results to those obtained elsewhere in the literature, finding good agreement especially for the results obtained from the Generation 2 ensembles, see Fig.~\ref{fig:kappa}, even though our ensembles have pions heavier than in nature and our results are not extrapolated to the continuum.
Nevertheless, it is intriguing that our method, based on hadronic correlation functions, gives similar results to the ones obtained with completely different observables, such as the renormalised chiral condensate and the strange quark number susceptibility.
%This suggests a universal nature of the chiral transition in QCD, because our method uses hadronic quantities and the other approaches use the renormalised chiral condensate and the strange quark number susceptibility.

In Section~\ref{sec:systematics} we assessed the systematic effects due to finite volume and our choice of $\tau_{\min}$. We find no sizeable finite-size effects, while an estimate for the systematic uncertainty coming from varying $\tau_{\min}$ is provided.

The lighter pion mass of the Generation 2L ensemble causes the correlation functions, especially the disconnected contributions, to be much noisier than for Generation 2, in particular near the pseudo-critical point.
The main contribution to the overall uncertainty of our results comes from the noisy disconnected contributions.
It will be interesting to study the effect of using a different type of noise, $e.g.$ $Z_2$ and $Z_4$ combined with different methods for noise reduction~\cite{Bali:2009hu,Giusti:2019kff}.

In the near future we plan to determine spectral functions to investigate the effects of chemical potential on the mesonic channels in order to obtain independent verification of our results.

\begin{acknowledgments}
We are grateful to the HadSpec collaboration for the use of their zero temperature ensembles.
This work is supported by the UKRI Science and Technology Facilities Council (STFC) Consolidated Grant No. ST/X000648/1.
We are grateful to Supercomputing Wales for the use of their computing resources
and to the Swansea Academy for Advanced Computing for support.
R.B. acknowledges support from a Science Foundation Ireland Frontiers for the Future Project award with grant number SFI-21/FFP-P/10186.
This work used the DiRAC Data Intensive service (DIaL2 / DIaL2.5) at the University of Leicester, managed by the University of Leicester Research Computing Service on behalf of the STFC DiRAC HPC Facility (www.dirac.ac.uk). The DiRAC service at Leicester was funded by BEIS, UKRI and STFC capital funding and STFC operations grants.
%DiRAC is part of the UKRI Digital Research Infrastructure,
It also used the DiRAC Blue Gene Q Shared Petaflop system at the University of Edinburgh, operated by the Edinburgh Parallel Computing Centre on behalf of the STFC DiRAC HPC Facility (www.dirac.ac.uk). This equipment was funded by BIS National E-infrastructure capital grant ST/K000411/1, STFC capital grant ST/H008845/1, and STFC DiRAC Operations grants ST/K005804/1 and ST/K005790/1. DiRAC is part of the National E-Infrastructure.
This work used the computing resources of the Irish Centre for High-End Computing (ICHEC).
This work was performed using the PRACE Marconi-KNL resources hosted by CINECA, Italy.
We acknowledge EuroHPC Joint Undertaking for awarding the project EHPC-EXT-2023E01-010 access to LUMI-C, Finland.
SK is supported by the National Research Foundation of Korea under grant NRF-2021R1A2C1092701 funded by the Korean government (MEST) and by the Institute of Information \& Communication Technology Planning \& Evaluation grant funded by the Korean government (Ministry of Science and ICT) (IITP-2024-RS-2024-00437191). 
\end{acknowledgments}

\noindent{\bf Authors' contributions}  --

Smecca: data analysis, physics interpretation, plot generation and primary manuscript production.

Allton: analysis idea, physics interpretations of results, and draft of the manuscript.

Aarts: contributions to analysis, physics interpretations of results, and manuscript.

Bignell, J\"ager, Skullerud: data production and contribution to the manuscript.

Nam, Kim, Wu: conceptual ideas and contribution to the manuscript.
 
\noindent
{\bf Research Data and Code Access} --
Gen 2 data can be found in Ref.~\cite{Gen2zenodo}. Gen 2L ensembles are available on request. Details of the code and data presented can be found in Refs.~\cite{mu_sq_data,mu_sq_workflow}.

\noindent
{\bf Open Access Statement} -- For the purpose of open access, the authors have applied a Creative Commons Attribution (CC BY) licence to any Author Accepted Manuscript version arising.

\appendix
\section{Lattice correlators}\label{app:Corrs}

In this appendix we show the correlators used to produce the main results in this paper and compare them with free correlators. In particular, we expand the discussion in Ref.~\cite{Nikolaev:2020vll}, where similar plots for the correlators are shown. The plots showing correlator ratios are new.

\begin{figure}[t]
  \centering
  \includegraphics[width=0.48\textwidth]{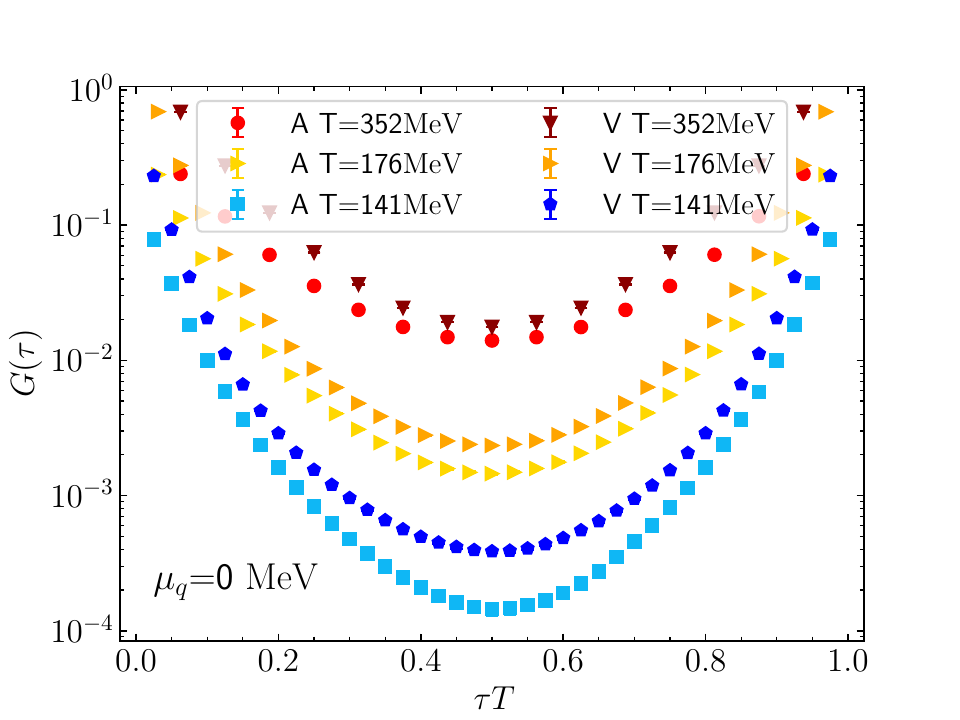}
  \includegraphics[width=0.48\textwidth]{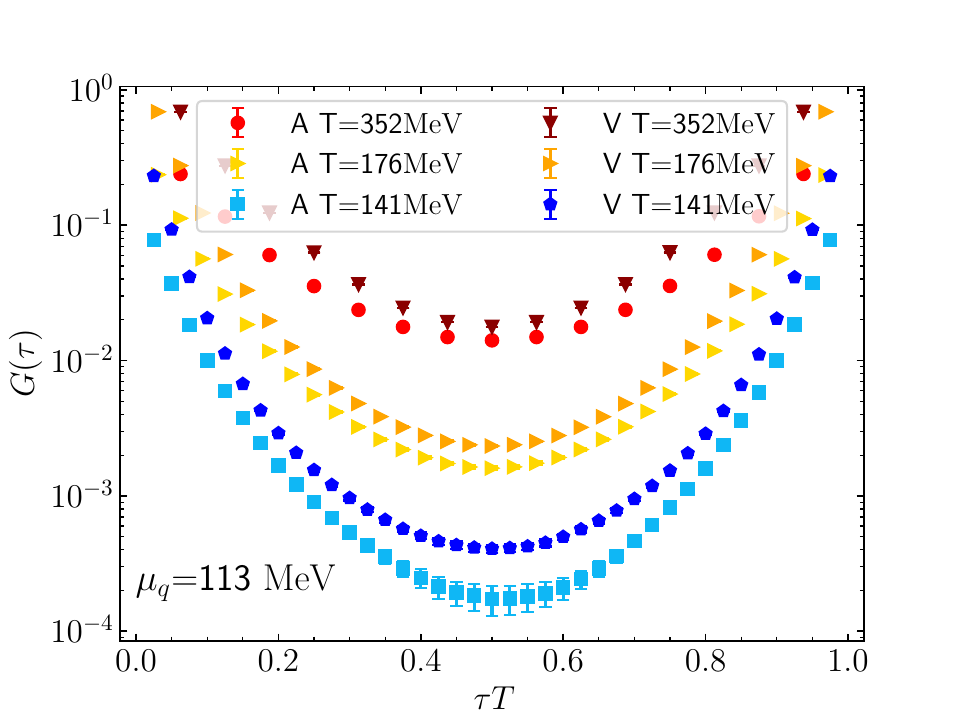}
  \caption{Generation 2 vector and axial-vector correlators, in lattice units, for $\mu_q=0$ (above) and $\mu_q=113$ $\mathrm{MeV}$ (below)
  at temperatures below ($T=141$ $\mathrm{MeV}$), near ($T=176$ $\mathrm{MeV}$) and above ($T=352$ $\mathrm{MeV}$) the pseudo-critical temperature obtained with the renormalised chiral condensate, $T_{\mathrm{pc}}=182(2)$ $\mathrm{MeV}$.}
  \label{fig:corrs}
\end{figure}

Fig.~\ref{fig:corrs} shows the vector and axial-vector correlators for selected temperatures and for two different values of $\mu_q$, in lattice units. As expected, for temperatures below $T_{\mathrm{pc}}$ the axial-vector correlator decreases exponentially faster than the vector one, indicating a heavier ground state. This behaviour is lost near $T_{\mathrm{pc}}$, where the two correlators look very similar. For temperatures above $T_{\mathrm{pc}}$ the correlators are closer near the midpoint. The effect of a non-zero chemical potential is hard to gauge from these plots, with the only visible effect shown by the axial-vector correlator below $T_{\mathrm{pc}}$ exhibiting a worse signal-to-noise ratio near the midpoint.

\begin{figure*}[t]
  \centering
  \includegraphics[width=.32\textwidth]{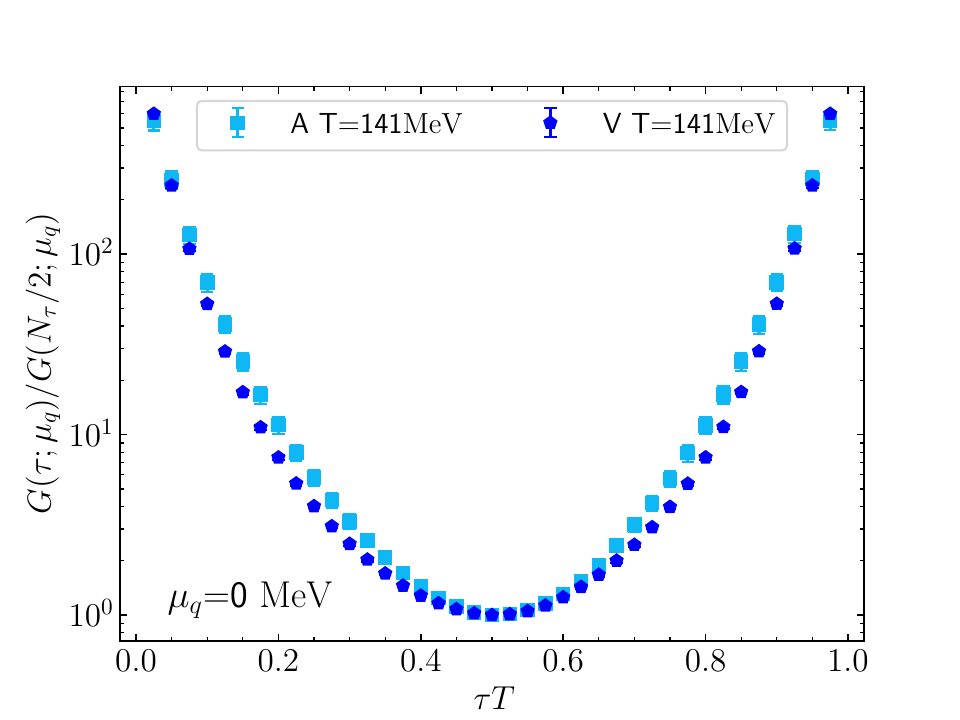}
  \includegraphics[width=.32\textwidth]{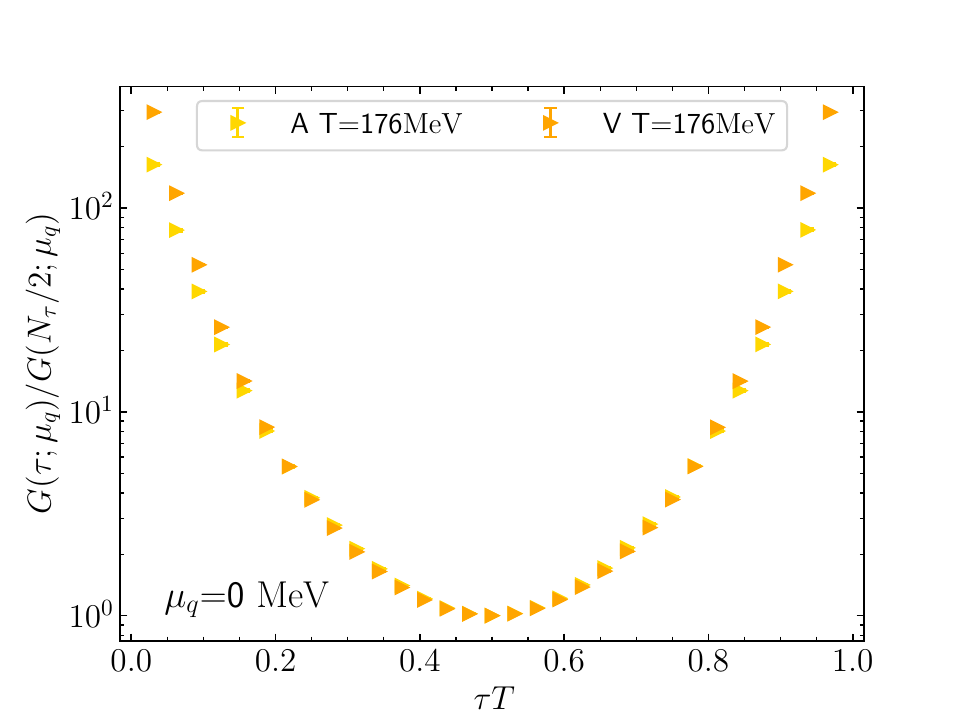}
  \includegraphics[width=.32\textwidth]{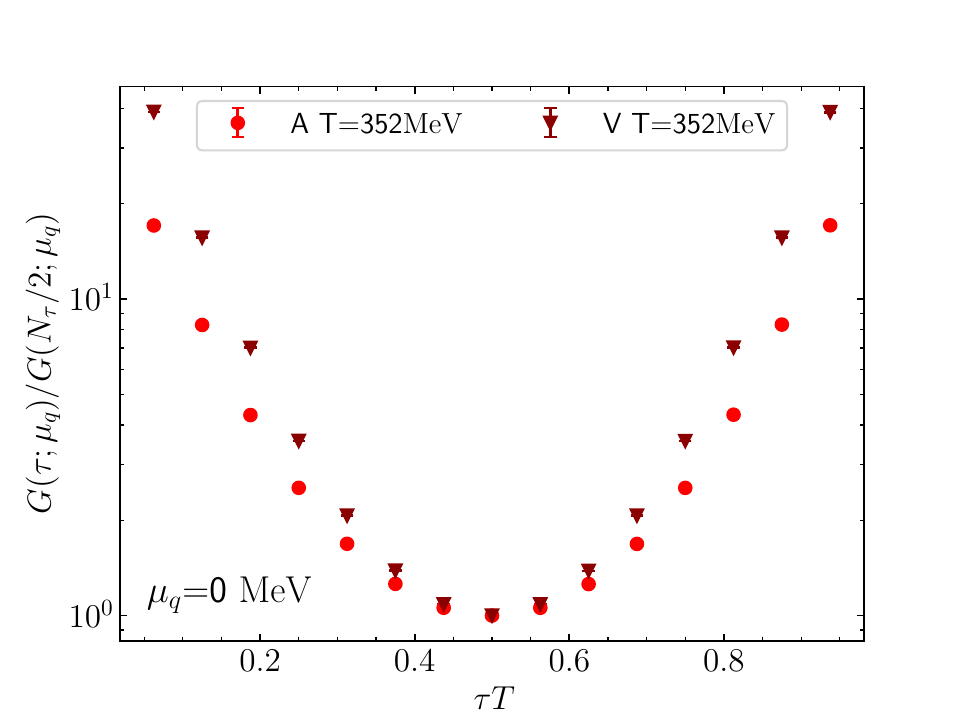}
  \includegraphics[width=.32\textwidth]{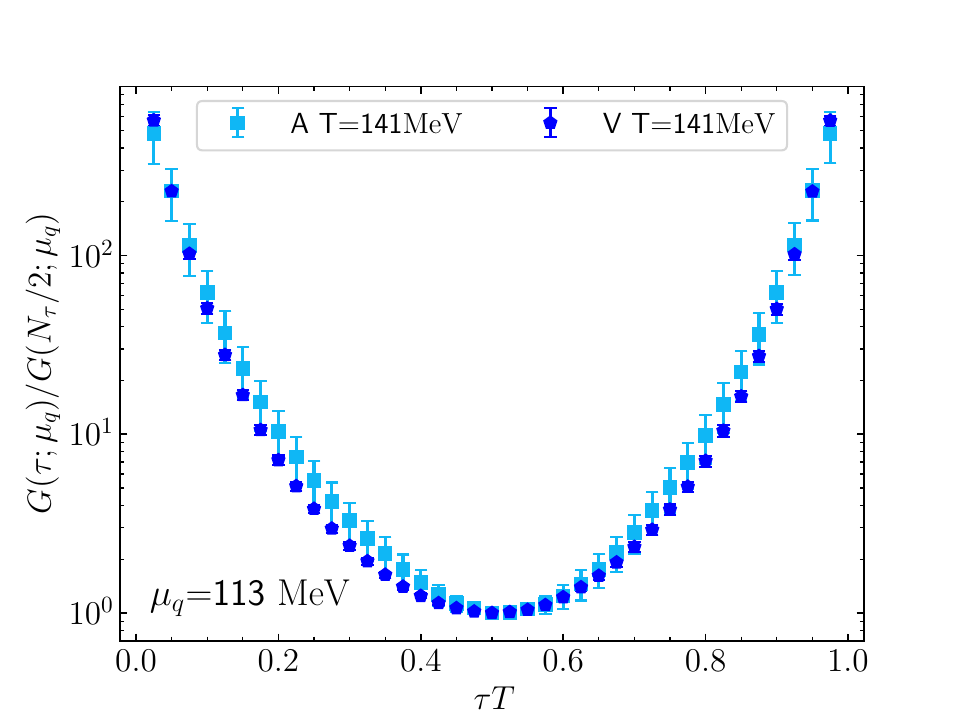}
  \includegraphics[width=.32\textwidth]{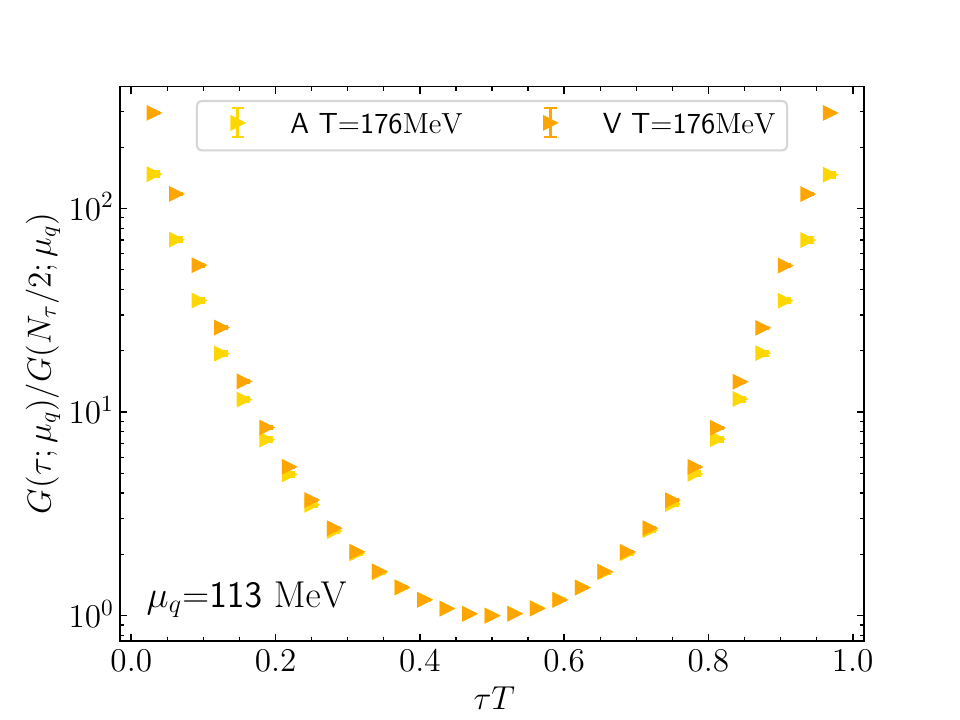}
  \includegraphics[width=.32\textwidth]{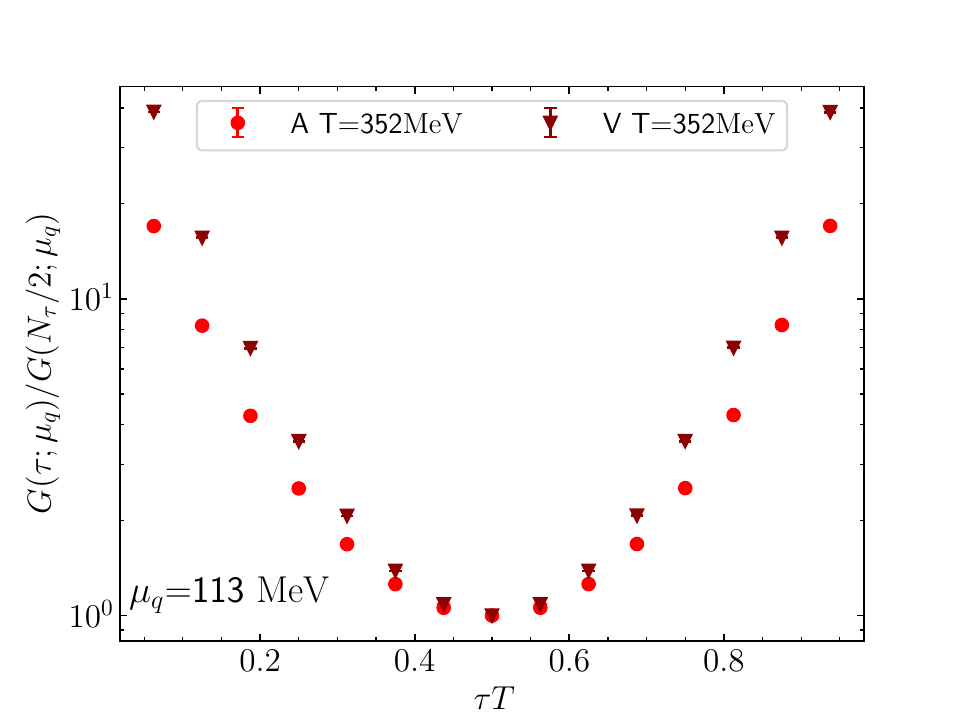}
  \caption{As in the preceding figure, with normalised correlators $G_{V/A}(\tau;\mu_q)/G_{V/A}(N_\tau/2;\mu_q)$.
  % Generation 2 vector and axial-vector correlators for $\mu_q=0$ (top row) and $\mu_q=113$ $\mathrm{MeV}$ (bottom row) at temperatures below ($T=141$ $\mathrm{MeV}$), near ($T=176$ $\mathrm{MeV}$) and above ($T=352$ $\mathrm{MeV}$) the pseudo-critical temperature obtained with the renormalised chiral condensate, $T_{\mathrm{pc}}=182(2)$ $\mathrm{MeV}$.
  }
  \label{fig:corrs_mu}
%\end{figure*}
%\begin{figure*}[h]
  \centering
  \includegraphics[width=0.48\textwidth]{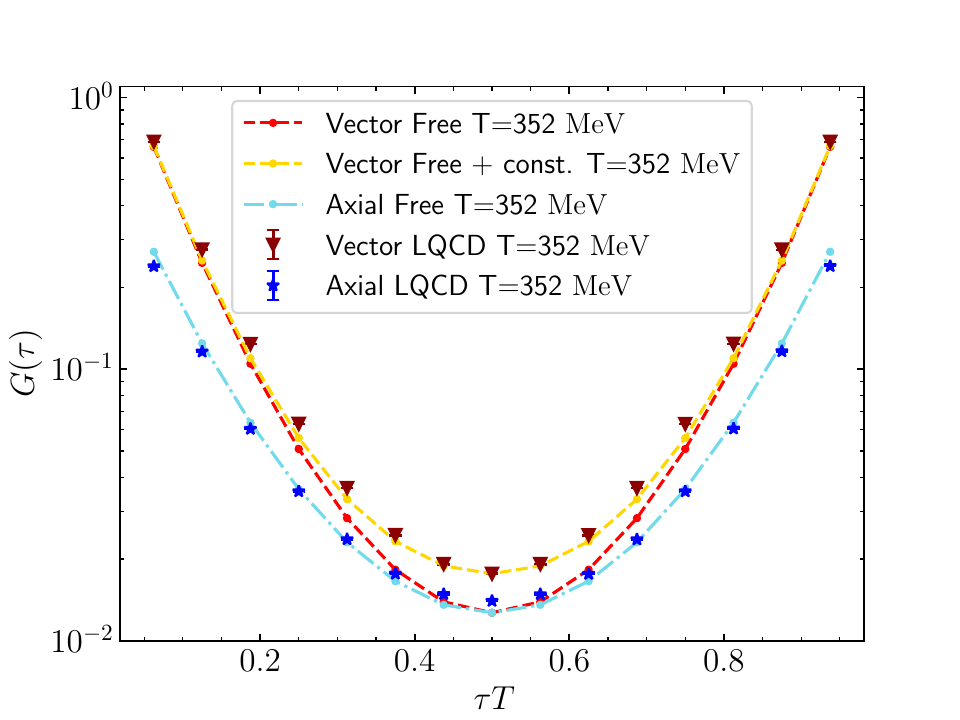}
  \includegraphics[width=0.48\textwidth]{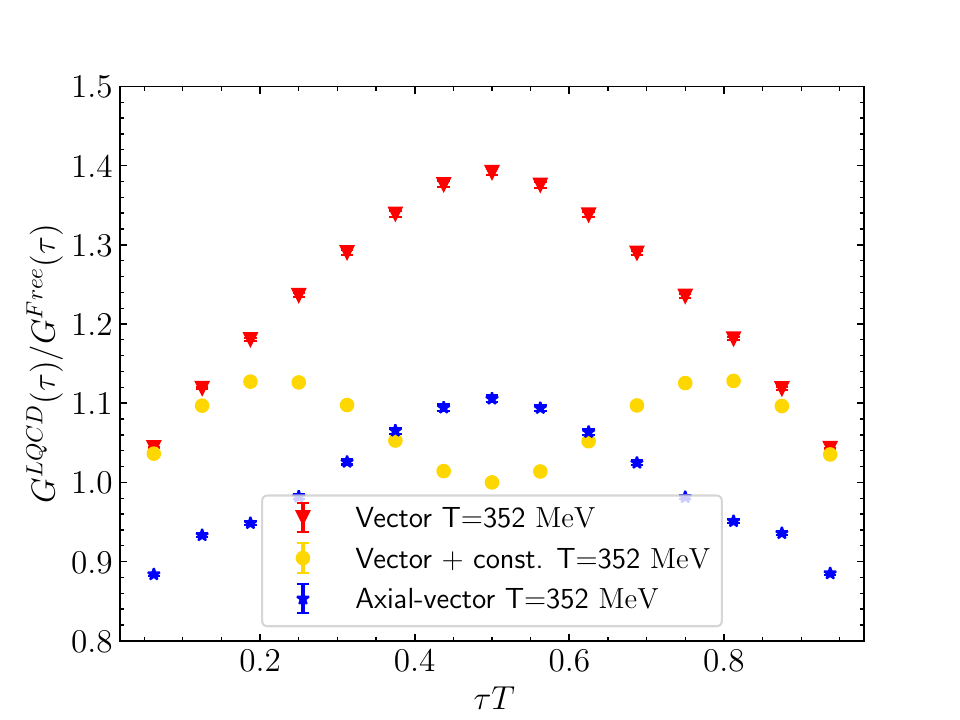}   
  \caption{Generation 2 vector and axial-vector correlators obtained in the full lattice QCD simulation and in the free theory, at $\mu_q=0$. Shown are correlators in lattice units (left) and ratio $G_{V/A}(\tau)/G^{\rm free}_{V/A}(\tau)$ (right). In the vector channnel, both the free correlator and the free correlator plus a constant, with const $= G_V^{\rm LQCD}(N_{\tau}/2)-G_V^{\rm free}(N_{\tau}/2)$, are shown.
  }
  \label{fig:free_vec}
\end{figure*}

The correlators used in this study are obtained using local currents with Wilson fermions and need to be renormalised.
To remove the dependence on the renormalisation when comparing the vector and axial-vector correlators, we normalise the correlators with respect to the temporal mid-point.
Fig.~\ref{fig:corrs_mu} shows the normalised vector and axial-vector correlators 
$G_{V/A}(\tau;\mu_q)/G_{V/A}(N_\tau/2;\mu_q)$ for selected temperatures and for two different values of $\mu_q$. 
Below $T_{\mathrm{pc}}$, the axial-vector correlator shows a stronger curvature compared to the vector correlator around the mid-point, consistent with a larger ground state mass.
Near $T_{\mathrm{pc}}$, the correlators are nearly degenerate, except near the edges where short-distance effects and lattice artifacts dominate.
At high temperature, lattice artifacts are even more pronounced due to the short temporal extent. Lattice artifacts affect the axial-vector and vector correlators differently as can be understood from the corresponding free correlators \cite{Aarts:2005hg}.
Increasing $\mu_q$ pushes the axial-vector correlator closer to the vector one while also introducing more noise. The effect of a non-zero $\mu_q$ is barely noticeable in Figs.~\ref{fig:corrs} and \ref{fig:corrs_mu}. In the main paper we use the ratio defined in Eq.~(\ref{eq:R-ratio}) to bring out the $T$ and $\mu_q$ dependence more prominently, see Fig.~\ref{fig:R-ratio}.

% \begin{figure}[ht!]
%   \centering
%   \includegraphics[width=0.48\textwidth]{figures/Correlators_mu_0.113_norm_Nt40.pdf}
%   \includegraphics[width=0.48\textwidth]{figures/Correlators_mu_0.113_norm_Nt32.pdf}
%   \includegraphics[width=0.48\textwidth]{figures/Correlators_mu_0.113_norm_Nt16.pdf}
%   \caption{Same as figure~\ref{fig:corrs_mu_zero} but for $\mu_q=113$ $\mathrm{MeV}$.}
%   \label{fig:corrs_mu}
% \end{figure}

Finally, in Fig.~\ref{fig:free_vec} we compare the correlators with the free correlators obtained using the same lattice geometry \cite{Aarts:2005hg}, in lattice units (left) and using the ratio $G_{V/A}(\tau)/G^{\rm free}_{V/A}(\tau)$ (right). Since the wave function normalisation is not included, there is an undetermined multiplicative factor. In the axial-vector channel, at this temperature good agreement between the full and the free correlator can be seen, with deviations up to around 10\%. In the vector channel, we observe a deviation between the full and the free correlator, which can be incorporated by adding a constant term to the free correlator. We determine this constant simply as the difference at the midpoint, $G_V^{\rm LQCD}(N_{\tau}/2)-G_V^{\rm free}(N_{\tau}/2)$, which improves the comparison. The origin of this shift may be due to the transport contribution. In the non-interacting case, the transport contribution to the spectral function in the vector channel reads \cite{Aarts:2020dda}
$\rho_{\rm transport}(\omega) = 2\pi N_c I \omega\delta(\omega)$,
where in the massless limit, $I=T^2/3$. This contribution yields a $\tau$-independent constant factor to the correlator, $G_{\rm transport}(\tau) = N_c I T$. 
The additional shift can then be attributed to a more pronounced transport peak, indicating that at this temperature, interaction effects remain present. It is worth noting that the electrical conductivity was studied in detail on the Generation 2 ensembles in Ref.~\cite{Aarts:2014nba} and hence we do not study it further here.

\bibliography{main}
\bibliographystyle{JHEP}

\end{document}